\pdfoutput=1
\documentclass[aps,prd,10pt,twocolumn,superscriptaddress]{revtex4-2}
\usepackage{amsmath,amssymb,amsfonts,array,tikz,xcolor,xcolor-material,hyperref}

\allowdisplaybreaks
\hypersetup{colorlinks}

\usetikzlibrary{arrows.meta,calc,decorations.markings}
\tikzset{
  x=1em, y=1em, node font=\footnotesize,
  ->-/.style = {
    decoration = {
      markings,
      mark = at position #1 with {\arrow{Stealth}},
    },
    postaction = {decorate},
  },
}

\newcommand{\I}{{\mathbb{I}}}
\newcommand{\ee}{{\mathrm{e}}}
\newcommand{\ii}{{\mathrm{i}}}
\DeclareMathOperator{\rank}{rank}

\newcommand{\Fusion}[3]{
  \begin{tikzpicture}[baseline=-0.5em]
    \draw [thick, MaterialGrey]
      (0:0) -- ( 30:1.2)
      (0:0) -- (150:1.2)
      (0:0) -- (270:1.2);
    \draw
      ( 1.2,1.3) node {#1}
      (-1.2,1.3) node {#2}
      ( 0, -1.8) node {#3};
  \end{tikzpicture}
}

\begin{document}

\title{Bulk Operator Reconstruction in Topological Tensor Network and\\ Generalized Free Fields}
\author{Xiangdong Zeng}
\affiliation{State Key Laboratory of Surface Physics, Fudan University, Shanghai 200433, China}
\affiliation{Department of Physics and Center for Field Theory and Particle Physics, Fudan University, Shanghai 200433, China}
\author{Ling-Yan Hung}
\email{lyhung@mail.tsinghua.edu.cn}
\affiliation{State Key Laboratory of Surface Physics, Fudan University, Shanghai 200433, China}
\affiliation{Department of Physics and Center for Field Theory and Particle Physics, Fudan University, Shanghai 200433, China}
\affiliation{Institute for Nanoelectronic devices and Quantum computing, Fudan University, Shanghai 200433, China}
\affiliation{Yau Mathematical Sciences Center, Tsinghua University, Beijing 100084, China}
\affiliation{Yanqi Lake Beijing Institute of Mathematical Sciences and Applications (BIMSA), Huairou District, Beijing 101408, China}
\date{\today}

\begin{abstract}
  In this paper, we would like to study operator reconstruction in a class of holographic tensor networks describing renormalization group flows studied in~\cite{chen2022exact}. We study examples of 2d bulk holographic tensor networks constructed from Dijkgraaf--Witten theories and found that for both $\mathbb{Z}_n$ group and $S_3$ group the number of bulk operators behaving like a generalized free field in the bulk scales as the order of the group. We also generalize our study to 3d bulks and found the same scaling for $\mathbb{Z}_n$ theories. However, there is no generalized free field when the bulk comes from more generic fusion categories such as the Fibonacci model.
\end{abstract}

\maketitle

\section{Introduction}

An important feature of the AdS/CFT correspondence~\cite{Maldacena:1997re} in the large $N$ limit is the emergence of free fields propagating in the AdS bulk, allowing a semi-classical description of the bulk theory, and computation of non-trivial quantities in the CFT via the bulk. By free fields, we mean that the correlation functions of these operators are computed approximately by Wick contraction. Generically, single trace operators in the CFT become approximately free Gaussian fields in the semi-classical dual bulk theory. This feature allows computations of correlation functions via Witten diagrams to be done in the AdS side~\cite{Witten:1998qj,Gubser:1998bc}. The emergence of generalized free fields in holographic CFT has also been discussed extensively. (See for example~\cite{Duetsch:2002hc,Liu:2018jhs,Collier:2018exn}, and very recently~\cite{Nebabu:2023iox} and references therein.)

Along a different vein, inspired by the Ryu--Taka\-yanagi formula~\cite{Ryu:2006bv}, it had been proposed that the tensor network is the appropriate framework to construct the linear map between the CFT degrees of freedom and the AdS ones~\cite{Swingle:2009bg}. The graphical representation of the appropriate tensor network looks like a discrete version of the AdS bulk. It describes a kind of coarse-graining process for boundary degrees of freedom, which also reside at the (asymptotic boundary) of the tensor network.

The bulk operators act on the auxiliary legs of the tensor network, and the CFT operators act on the legs located at the asymptotic boundary of the network. Clearly, the tensor network is providing a linear map between these operators~\cite{Pastawski:2015qua,Hayden:2016cfa}.

Due to the fact that the tensor network is local -- i.e.\ it is decomposed into a product of tensors that only contract with some number of neighboring tensors (the precise number depends on the dimension of the bulk space), the bulk boundary map can be read off via ``operator pushing''~\cite{Pastawski:2015qua}. (Examples are also discussed extensively in~\cite{Bhattacharyya:2016hbx,Bhattacharyya:2017aly}.)

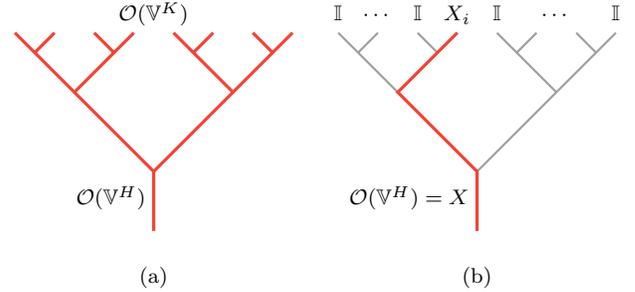
\begin{figure}[htb]
  \centering
  \def\a{0.75}
  \def\b{0.75}
  \def\PathI{
    (0,0) -- (0,-3*\b)
    (-7*\a,7*\b) -- (0,0) -- (7*\a,7*\b)
    (-4*\a,4*\b) -- (-\a,7*\b)
    ( 4*\a,4*\b) -- ( \a,7*\b)
    (-6*\a,6*\b) -- (-5*\a,7*\b)
    ( 6*\a,6*\b) -- ( 5*\a,7*\b)
    (-2*\a,6*\b) -- (-3*\a,7*\b)
    ( 2*\a,6*\b) -- ( 3*\a,7*\b)
  }
  \begin{tikzpicture}
    \draw [very thick, MaterialRed] \PathI;
    \draw (0,7*\b) node [above=1.2ex, anchor=base] {$\mathcal{O}(\mathbb{V}^K)$}
          (0,-1) node [left] {$\mathcal{O}(\mathbb{V}^H)$}
          (0,-4) node {(a)};
  \end{tikzpicture}
  \quad
  \begin{tikzpicture}
    \draw [thick, MaterialGrey] \PathI;
    \draw [very thick, MaterialRed] (0,-3*\b) -- (0,0) -- (-4*\a,4*\b) -- (-\a,7*\b);
    \draw foreach \i in {-7, -3, 1, 7} {(\i*\a,7*\b) node [above=1.2ex, anchor=base] {$\I$}}
          foreach \i in {-5, 4}        {(\i*\a,7*\b) node [above=1.2ex, anchor=base] {$\cdots$}}
          (-\a,7*\b) node [above=1.2ex, anchor=base] {$X_i$}
          (0,-1) node [left] {$\mathcal{O}(\mathbb{V}^H)=X$}
          (0,-4) node {(b)};
  \end{tikzpicture}
  \caption{Operator pushing in tree-like tensor networks. (a) Bulk operator $\mathcal{O}(\mathbb{V}^H)$ is reconstruced by boundary operator $\mathcal{O}(\mathbb{V}^K)$, where we take $H=1$ for simplicity. (b) Bulk operator $\mathcal{O}(\mathbb{V}^H)=X$ is reconstruced by a simple form boundary operator $\mathcal{O}(\mathbb{V}^K)=\sum_i\alpha_i(\I_1\otimes\cdots\otimes X_i\otimes\dots\otimes\I_{K})$ as in Equation~\eqref{eq:operator-pushing-coefficients}. Hence it corresponds to a generalized free field.}
  \label{fig:operator-pushing}
\end{figure}

This leads us to a natural question: what kind of tensor networks most resemble the bulk-boundary map encountered in AdS/CFT where the bulk is describable by a weakly coupled bulk theory where the bulk fields are almost free? This question was addressed in~\cite{Bhattacharyya:2017aly}. Bulk operators act on internal legs in the bulk. Under operator pushing, its action on the tensor is equivalent to another set of operators acting on some other legs. This is illustrated in \autoref{fig:operator-pushing}. In such a reconstruction one has to make a choice over which of the legs are in-coming and which are out-going. There is a canonical choice when the tensor network is essentially performing coarse graining, where operators acting on degrees of freedom after coarse graining should be related to operators acting on the pre-coarse grained degrees of freedom. In this case, it is shown~\cite{Bhattacharyya:2017aly} that a generalized free field should be such that when they are pushed across a tensor, they can be decomposed as a sum of ``simple operators''. To be precise, if the tensor network is made up of a network of coarse-graining tensor $M^{i_1 \cdot i_K}_{j_1 \cdot j_H}$ that takes $\mathbb{V}^K \to \mathbb{V}^H$ (where $K>H$, $i_i, j_i \in \mathbb{V}$, and $\mathbb{V}$ is a $d$-dimensional vector space) then operator pushing corresponds to
\begin{equation}
  \mathcal{O} \bigl( \mathbb{V}^H \bigr) \cdot M = M \cdot \mathcal{O} \bigl( \mathbb{V}^K \bigr).
  \label{eq:operator-pushing}
\end{equation}
An almost free bulk operator acting on one of the legs among $\mathbb{V}^H$, i.e.
\begin{equation}
    \mathcal{O} \bigl( \mathbb{V}^H \bigr)
  = \I_1 \otimes \cdots \otimes \I_{i-1} \otimes X_i \otimes \I_{i+1} \otimes \dots \otimes \I_{H},
\end{equation}
and it satisfies Equation~\eqref{eq:operator-pushing}, such that
\begin{equation}
    \mathcal{O} \bigl( \mathbb{V}^K \bigr)
  = \sum_{i=1}^K \alpha_i \, \bigl(
      \I_1 \otimes \cdots \otimes \I_{i-1} \otimes X_i \otimes \I_{i+1} \otimes \dots \otimes \I_{K}
    \bigr),
  \label{eq:operator-pushing-coefficients}
\end{equation}
for some constants $\{\alpha_i\}$. When this is satisfied, the reconstruction of the bulk operator $\mathcal{O}(l_B) $ acting on a bulk leg $l_B$ in terms of boundary operators would take the form of
\begin{equation}
  \mathcal{O}(l_B) = \sum_b K^I (l_b, l_B)  \mathcal{O}^I(l_b),
\end{equation}
where $\{\mathcal{O}^I\}$ is a complete set of basis operators acting on each leg $l_b$ at the boundary, and $K^I(l_b, l_B)$ is the bulk-boundary kernel which can be expressed in terms of the $\alpha_i$ in Equation~\eqref{eq:operator-pushing-coefficients}. This expression has the same form as the HKLL kernel constructed from bulk boundary propagators~\cite{Hamilton:2005ju,Hamilton:2006az}, and one can show that correlation functions of bulk operators would behave like generalized free fields~\cite{Bhattacharyya:2017aly,Hung:2019zsk}. We note that since $T$ is a rectangular matrix, the reconstruction is not unique. However, a generalized free field would be one where Equation~\eqref{eq:operator-pushing-coefficients} can be satisfied at all.

In this paper, we would like to explore families of coarse-graining tensor networks that follow from topological theories, introduced and discussed in~\cite{Bal:2018wbw,chen2022exact}. They are interesting because they are key to recovering families of CFTs, and the coarse graining or RG tensors carry resemblances to the AdS bulk which is checked numerically at least in low dimensions. It is thus interesting to study operator pushing in these RG tensor networks and explore when generalized free fields might emerge.

We would begin our analysis with RG operators constructed from 1+1d topological field theory. In particular, we would focus on the trivial class of Dijkgraaf--Witten theories with gauge group $G$. It is found that for $G=\mathbb{Z}_n$ there would be $n$ generalized free fields. We also study the simplest non-abelian theory with $G=S_3$.

This study is generalized to RG operators constructed from 2+1d topological field theory. We study both the Dijkgraaf--Witten type lattice gauge theories, and also simple examples of Turaev--Viro type theories. In the case of $\mathbb{Z}_n$ lattice gauge theories, it is found that the number of generalized free fields is given by $n$. We also compute examples of operator reconstruction in the stereotypical example of topological orders in 2+1 dimensions, namely the Fibonacci model. In this case, we found no generalized free operator at all. 

\section{Operator pushing in 1+1d}

We begin with the 1+1d Dijkgraph--Witten theory characterized by group $G$. An RG operator can be constructed from the topological theory~\cite{Bal:2018wbw,chen2022exact}, which takes the form of a tree network. Each vertex is 3-valent, and for the untwisted version of the Dijkgraaf--Witten theory (which is the main focus of the present section), each 3-valent vertex resides a 3-index tensor $M^{g_1, g_2}_{g_3}$ that takes the following form:
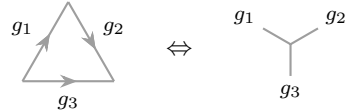
\[
  \begin{tikzpicture}[baseline=0pt]
    \draw [thick, MaterialGrey, ->-=0.3, ->-=0.8]
      (210:2) -- ( 90:2)
      ( 90:2) -- (-30:2);
    \draw [thick, MaterialGrey, ->-=0.6]
      (210:2) -- (-30:2);
    \draw
      ( 30:2) node {$g_2$}
      (150:2) node {$g_1$}
      (270:1.8) node {$g_3$};
  \end{tikzpicture}
  \quad \Leftrightarrow \quad
  \begin{tikzpicture}[baseline=-4pt]
    \draw [thick, MaterialGrey]
      (0:0) -- ( 30:1.2)
      (0:0) -- (150:1.2)
      (0:0) -- (270:1.2);
    \draw
      ( 30:2) node {$g_2$}
      (150:2) node {$g_1$}
      (270:1.8) node {$g_3$};
  \end{tikzpicture}
\]
where $g_1$, $g_2$ and $g_3$ are elements of $G$. The tensor $M$ imposes the group product or fusion rule of $G$, such that it is non-vanishing only for $G(g_1,g_2) \equiv g_1\times g_2=g_3$. i.e.
\begin{equation}
  M^{ij}_{ k} = \delta_{G(i,j), k}.
\end{equation}

\begin{figure}[htb]
  \centering
  \begin{tikzpicture}
    \begin{scope}
      \def\r{3.5}
      \draw [thick, MaterialGrey]
        foreach \x/\xText in {20/e,55/d,90/c,125/b,160/a} {
          (0,0) -- (\x:\r) (\x:1.2*\r) node [black] {$\xText$}
        }
        (0,0) node [below, black] {$o$} circle [radius=\r];
      \draw [very thick, MaterialRed]  (20:\r) -- (90:\r) -- (160:\r);
      \draw [very thick, MaterialBlue] (20:\r) -- (160:\r);
      \draw (195:0.6*\r) node [rotate= 70] {$\cdots$}
            (-15:0.6*\r) node [rotate=-70] {$\cdots$}
            (6,0) node {$\implies$};
    \end{scope}
    \begin{scope}[xshift=12em]
      \def\a{1}
      \def\b{1}
      \draw [thick, MaterialGrey]
        (0,0) -- (0,-2*\b)
        (-3*\a,3*\b) -- (0,0) -- (3*\a,3*\b)
        (-1.8*\a,1.8*\b) -- (-0.6*\a,3*\b)
        ( 1.8*\a,1.8*\b) -- ( 0.6*\a,3*\b);
      \draw [very thick, MaterialRed]
        (-3.5*\a,1.5*\b) -- (-1.8*\a,2.6*\b) -- (-0.1*\a,1.5*\b) -- cycle
        ( 3.5*\a,1.5*\b) -- ( 1.8*\a,2.6*\b) -- ( 0.1*\a,1.5*\b) -- cycle;
      \draw [very thick, MaterialBlue]
        (-1.7*\a,-0.4*\b) -- (0,0.8*\b) -- (1.7*\a,-0.4*\b) -- cycle;
      \draw ( 6.8*\a,1.5*\b) node {$\left.\vphantom{\rule{0.1em}{5ex}}\right\}\mathcal{U}(G,\alpha)$}
            (-4*\a,2.5*\b) node {$\cdots$}
            ( 4*\a,2.5*\b) node {$\cdots$}
            (   0,-2.5*\b) node {$\vdots$};
    \end{scope}
    \begin{scope}[xshift=3.5em, yshift=-8.5em]
      \def\a{1.5}
      \def\b{2}
      \draw [thick, MaterialGrey]
        (\a,0) -- (0,\b) -- (-\a,0) -- (0,-\b) -- cycle
        (-\a,0) -- (\a,0);
      \draw [thick, MaterialGrey, xshift=4*\a em]
        (\a,0) -- (0,\b) -- (-\a,0) -- (0,-\b) -- cycle
        (0,-\b) -- (0,\b);
      \draw (1.9*\a,0) node {=};
    \end{scope}
    \draw (6,-4)  node {(a)}
          (6,-11) node {(b)};
  \end{tikzpicture}
  \caption{(a) The triangulation of the disk can be converted to a tree-like tensor network by repeated use of the associativity condition $\alpha(g_1,g_2)\alpha(g_1 g_2,g_3)=\alpha(g_1,g_2 g_3)\alpha(g_2,g_3)$ which is also illustrated pictorially in (b). Here, each triangle is assigned a value $\alpha(g_i,g_j)\in H^2(G,U(1))$ where $H^2$ denotes the 2-cohomology. The boundary circle with $2N$ edges is converted to another one with $N$ edges by using this condition. The collection of triangles connecting the two circles gives the RG operator $\mathcal{U}(G,\alpha)$. Images from~\cite{chen2022exact} (with modification).}
  \label{fig:rg-1+1d}
\end{figure}
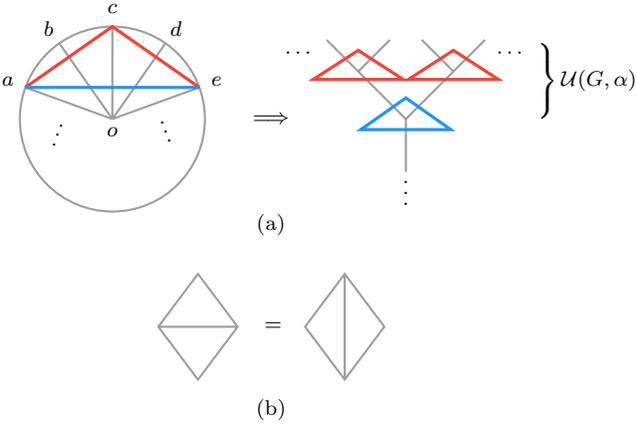

The tensor network has a 1d boundary (see \autoref{fig:rg-1+1d}). When inserting an operator $B$ in the bulk, its action is equivalent to some other operators at the boundary. Finding the boundary operator that recreates the action of a given bulk operator is the problem of bulk operator reconstruction. Since the tensor network is local, we can reconstruct the action of the bulk operator by studying the reconstruction of the bulk operator across one constituent tensor in the RG tensor network. Specifically, reconstruction or operator pushing across one constituent tensor amounts to finding an operator $A$ for given operator $B$, such that
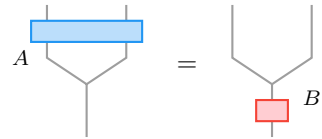
\[
  \begin{tikzpicture}[baseline=2.5em]
    \draw [thick, MaterialGrey]
      (1.5,5) -- (1.5,3) -- (0,2) -- (-1.5,3) -- (-1.5,5)
      (0,2) -- (0,0);
    \draw [thick, MaterialBlue, fill=MaterialBlue100]
      ( 2.1,3.6) -- ( 2.1,4.4) -- (-2.1,4.4) -- (-2.1,3.6) -- cycle;
    \draw (-2.5,3) node {$A$};
  \end{tikzpicture}
  \quad = \quad
  \begin{tikzpicture}[baseline=2.5em]
    \draw [thick, MaterialGrey]
      (1.5,5) -- (1.5,3) -- (0,2) -- (-1.5,3) -- (-1.5,5)
      (0,2) -- (0,0);
    \draw [thick, MaterialRed, fill=MaterialRed100]
      ( 0.6,0.6) -- ( 0.6,1.4) -- (-0.6,1.4) -- (-0.6,0.6) -- cycle;
    \draw (1.5,1.5) node {$B$};
  \end{tikzpicture}
\]
In the case where the bulk operator $B$ is a ``generalized free field'', the reconstruction by $A$ should be expressible in a ``simple'' form as discussed above (i.e.\ $A$ is a generic linear combination of $\I\otimes\tilde{A}$ and $\tilde{A}\otimes\I$). To ensure that the operator would behave like a generalized free field, we would require that $\tilde{A}$ falls back into the set of operators $\{B\}$ that admits simple reconstruction.

The general reconstruction equation for one con\-stit\-u\-ent tensor is given by
\begin{equation}
  A_{(ij), (i'j')} M^{i'j'}_{ k} = M^{ij}_{k'} B_{k'k}.
  \label{eq:1+1d-constraint}
\end{equation}
In the following, we would consider solving Equation~\eqref{eq:1+1d-constraint} for generic operator $B$, and also identifying the set of generalized free fields from it. We note that the collection of $B$ whose reconstruction $A$ that is ``simple'' forms a complete basis of generalized free fields in the tensor network.

The vector space residing on each leg of a tensor would generically be of dimension $|G|$. We can construct a basis for the operators acting on each leg using the generalized Pauli matrices~\cite{patera1988pauli}. For a finite group $G$ with $|G|=n$ and elements labeled by $0,1,\dots,n-1$, one can construct basis states for the $|G|$ dimensional vector space at each leg, given by
\begin{equation}
  |i\rangle = \begin{pmatrix} 0 \\ \vdots \\ 1_{\text{$i$-th}} \\ \vdots \\ 0 \end{pmatrix}, \quad
  i \in \{ 0, 1, \dots, n-1 \}.
\end{equation}
In such basis, the generalized Pauli matrices are generated by \emph{shift matrix} $X$ and \emph{clock matrix} $Z$ as follows:
\begin{equation}
  X = \begin{pmatrix}
    0 & 1 & 0 & \cdots & 0 \\
    0 & 0 & 1 & \cdots & 0 \\
    \vdots & \vdots & \vdots & \ddots & \vdots \\
    0 & 0 & 0 & \cdots & 1 \\
    1 & 0 & 0 & \cdots & 0
  \end{pmatrix}, \quad
  Z = \begin{pmatrix}
    1 & 0 & 0 & \cdots & 0 \\
    0 & \omega & 0 & \cdots & 0 \\
    \vdots & \vdots & \vdots & \ddots & \vdots \\
    0 & 0 & 0 & \cdots & \omega^{n-1}
  \end{pmatrix},
  \label{eq:generalized-pauli-matrices}
\end{equation}
where $\omega=\ee^{2\pi\ii/n}$ is the $n$-th root of unity. Then the generalized Pauli matrices are
\begin{equation}
  \sigma_\mu := \sigma_{ns+t} = X^t Z^s,
\end{equation}
where $s=\lfloor\mu/n\rfloor$, $t=\mu\bmod n$ and $\mu\in\{0,1,\dots,n^2-1\}$. Now Equation~\eqref{eq:1+1d-constraint} becomes
\begin{align}
  &\mathrel{\phantom{=}}
     A_{(ij), (i'j')} \delta_{G(i',j'), k} \notag \\
  &= \delta_{G(i,j), k'} B_{k'k}
   = \delta_{G(i,j), k'} (\sigma_\mu)_{k'k}
   = (\sigma_\mu)_{G(i,j), k}.
  \label{eq:1+1d-constraint-in-sigma}
\end{align}
So once $G(i,j)$ is specified, we can solve the above equation for each $\sigma_\mu$ via standard linear algebra methods. A solution is given by
\begin{equation}
  A^{(\mu)}_{(ij), (i'j')} = \begin{cases}
    (\sigma_\mu)_{G(i,j), j'}, & i' = 0; \\
    0. & i' \neq 0.
  \end{cases}
  \label{eq:1+1d-specific-solution}
\end{equation}
As expected since $M$ is a rectangular matrix, the solution above is only determined up to the addition of any linear combination of the collection of homogeneous solutions satisfying $A_{(ij),(i'j')}\delta_{G(i',j'),k}=0$

We would like to look for the subset of bulk operators such that there exists $A$ that is a local/simple operator such that it admits non-trivial action only on one leg, and acts trivially on the rest, i.e.
\begin{equation}
  A_{(ij), (i'j')} = \tilde{A}_{ii'} \delta_{jj'} \enspace \text{or} \enspace
  A_{(ij), (i'j')} = \delta_{ii'} \tilde{A}_{jj'}.
\end{equation}
Then
\begin{align}
  \tilde{A}_{ii'} \delta_{G(i',j), k} &= (\sigma_\mu)_{G(i,j), k} \notag \\
  \text{or} \enspace
  \tilde{A}_{jj'} \delta_{G(i,j'), k} &= (\sigma_\mu)_{G(i,j), k}, \notag \\
  &\quad \forall j \in \{ 0, 1, \dots, n-1 \}.
  \label{eq:1+1d-simple-form-constraint}
\end{align}
$\tilde{A}$ can be easily solved too. The necessary and sufficient condition that the above equation has a solution is
that for each column of $((\sigma_\mu)_{G(i,j),k})^{\mathrm{T}}_s$, the rank of the augmented matrix (i.e.\ appending this column vector to the matrix) satisfies
\begin{align}
  &\mathrel{\phantom{=}}
     \rank\bigl[ (\delta_{G(i',j), k})^{\mathrm{T}} \big| \bigl( (\sigma_\mu)_{G(i,j),k} \bigr)^{\mathrm{T}}_s \bigr] \notag \\
  &= \rank\bigl[ (\delta_{G(i',j), k})^{\mathrm{T}} \bigr]
   = \rank\bigl( \delta_{G(i',j), k} \bigr) = n.
\end{align}
Practically, we can take coefficients $\{\alpha_\mu\}$ such that
\begin{align}
  \tilde{A}_{ii'} \delta_{G(i',j), k} - \sum_\mu \alpha_\mu (\sigma_\mu)_{G(i,j), k} &= 0 \notag \\
  \text{or} \enspace
  \tilde{A}_{jj'} \delta_{G(i,j'), k} - \sum_\mu \alpha_\mu (\sigma_\mu)_{G(i,j), k} &= 0,
  \label{eq:1+1d-simple-form-augmented-constraint}
\end{align}
which is a set of homogeneous equations with respect to $\tilde{A}_{ii'}$ (or $\tilde{A}_{jj'}$) and $\alpha_\mu$. The coefficient matrix $\tilde{M}$ will then give $B=\sum_\mu\alpha_\mu\sigma_\mu$ and the corresponding $A$ operator.

In the following, we will present some concrete results for special classes of theories.

\subsection{Abelian case: \texorpdfstring{$\mathbb{Z}_n$}{ℤₙ} group}

\subsubsection{General solution}

For $\mathbb{Z}_n$ group, the fusion rules are given by modular arithmetic:
\begin{equation}
  G(i,j) = (i+j)\bmod n,
  \label{eq:zn-fusion-rules}
\end{equation}
Then $\delta_{G(i,j),k}=\delta_{(i+j)\bmod n,k}$ is a rank-$n$ matrix, whose (transposed) null space is spanned by $n^2-n$ vectors $v^{(p)}$ such that:
\begin{multline}
  v^{(p)}_q = \delta_{(-p-\lfloor p/n\rfloor-2)\bmod n, q} - \delta_{n^2-p-1, q}, \\
  p \in \left\{ 0, \dots, n^2-n-1 \right\}, \,
  q \in \left\{ 0, \dots, n^2-1 \right\}.
\end{multline}
The general solution $A^*$ is given by the linear combination of $v^{(p)}$:
\begin{equation}
  A^* = \begin{pmatrix}
    \beta_{0,0} v^{(0)} + \dots + \beta_{0,n^2-n-1} v^{(n^2-n-1)} \\
    \vdots \\
    \beta_{n^2-1,0} v^{(0)} + \dots + \beta_{n^2-1,n^2-n-1} v^{(n^2-n-1)}
  \end{pmatrix}.
\end{equation}
where $\beta_{ij}$ are arbitrary constants. The specific solution part is
\begin{equation}
  A^{(\mu)}_{(ij), (0j')} = (\sigma_\mu)_{(i+j)\bmod n, j'},
\end{equation}
as we have already mentioned in Equation~\eqref{eq:1+1d-specific-solution}.

To solve for bulk operators with local/simple reconstruction, it can be seen that the coefficient matrix $\tilde{M}$ in Equation~\eqref{eq:1+1d-simple-form-augmented-constraint} has dimension $n^3\times2n^2$ and rank $2n^2-n$. The solution is
\begin{equation}
  \tilde{A}^{(k)}_{ii'} = (\sigma_k)_{ii'}, \quad
  \alpha^{(k)}_\mu = \delta_{k\mu}
\end{equation}
where $k\in\{0,1,\dots,n-1\}$. Thus it means that the generalized free bulk operator given by $B=\sigma_k$ corresponds to a simple operator
\begin{equation}
  A = \I \otimes \sigma_k \enspace \text{or} \enspace \sigma_k\otimes\I.
  \label{eq:zn-solution}
\end{equation}
Since $\tilde{A}=\sigma_k$ falls in the set of $B$, this operator pushing procedure can be further iterated.

\subsubsection{Example: \texorpdfstring{$\mathbb{Z}_2$}{ℤ₂}}

Let's take $\mathbb{Z}_2$ group ($n=2$) as an explicit example. The null space of $M^{\mathrm{T}}=\delta_{k,i+j\bmod2}$ is spanned by
\begin{equation}
  \{ v^{(p)} \} = \Biggl\{
    \begin{pmatrix} 1 \\ 0 \\ 0 \\ -1 \end{pmatrix}, \,
    \begin{pmatrix} 0 \\ 1 \\ -1 \\ 0 \end{pmatrix}
  \Biggr\},
\end{equation}
which is equivalent to
\begin{equation}
  A^* = \begin{pmatrix}
    \beta_{0,0} & \beta_{0,1} & -\beta_{0,1} & -\beta_{0,0} \\
    \beta_{1,0} & \beta_{1,1} & -\beta_{1,1} & -\beta_{1,0} \\
    \beta_{2,0} & \beta_{2,1} & -\beta_{2,1} & -\beta_{2,0} \\
    \beta_{3,0} & \beta_{3,1} & -\beta_{3,1} & -\beta_{3,0}
  \end{pmatrix},
\end{equation}
The full solutions are then
\begin{equation}
  \begin{aligned}
    B = \sigma_0 = \I           &\implies A = A^* + \begin{pmatrix} 1 &  0 & 0 & 0 \\ 0 &  1 & 0 & 0 \\ 0 &  1 & 0 & 0 \\ 1 &  0 & 0 & 0 \end{pmatrix}, \\
    B = \sigma_1 = \sigma_x     &\implies A = A^* + \begin{pmatrix} 0 &  1 & 0 & 0 \\ 1 &  0 & 0 & 0 \\ 1 &  0 & 0 & 0 \\ 0 &  1 & 0 & 0 \end{pmatrix}, \\
    B = \sigma_2 = \sigma_z     &\implies A = A^* + \begin{pmatrix} 1 &  0 & 0 & 0 \\ 0 & -1 & 0 & 0 \\ 0 & -1 & 0 & 0 \\ 1 &  0 & 0 & 0 \end{pmatrix}, \\
    B = \sigma_3 = -\ii\sigma_y &\implies A = A^* + \begin{pmatrix} 0 & -1 & 0 & 0 \\ 1 &  0 & 0 & 0 \\ 1 &  0 & 0 & 0 \\ 0 & -1 & 0 & 0 \end{pmatrix}.
  \end{aligned}
  \label{eq:z2-solution}
\end{equation}
We can see that only $\sigma_x$ inserted in bulk corresponds to a simple form
\begin{equation}
  A = \I \otimes \sigma_x = \begin{pmatrix}
    0 & 1 & 0 & 0 \\
    1 & 0 & 0 & 0 \\
    0 & 0 & 0 & 1 \\
    0 & 0 & 1 & 0
  \end{pmatrix}
\end{equation}
or
\begin{equation}
  A = \sigma_x \otimes \I = \begin{pmatrix}
    0 & 0 & 1 & 0 \\
    0 & 0 & 0 & 1 \\
    1 & 0 & 0 & 0 \\
    0 & 1 & 0 & 0
  \end{pmatrix}
\end{equation}
at the boundary, which is consistent with Equation~\eqref{eq:zn-solution}.

\subsubsection{Higher level trees}

When considering the full holographic network that is a tree with many layers of the constituent tensors studied above (e.g. with total number of layers $L$, see \autoref{fig:rg-1+1d}), we can still find a solution by iteratively using the above method. When $B=\sigma_k$ where $k\in\{0,1,\dots,n-1\}$, since
\begin{equation}
  A_1 = \sigma_k \otimes \I \enspace \text{or} \enspace \I \otimes \sigma_k,
\end{equation}
the boundary operator at level $L$ is still simple and can be written as
\begin{equation}
  A_L = \I^{\otimes L-l} \otimes \sigma_k \otimes \I^{\otimes l-1}, \quad l = 0,1,\dots,L.
\end{equation}
Any other generalized free bulk operators can be constructed from linear combinations of these $B$ operators.

\subsection{Non-abelian case: \texorpdfstring{$S_3$}{𝑆₃} group}

The group multiplication table for $S_3$ is
\begin{center}
  \begin{tabular}{c|cccccc}
    & $g_0$ & $g_1$ & $g_2$ & $g_3$ & $g_4$ & $g_5$ \\
    \hline
    $g_0$ & 0 & 1 & 2 & 3 & 4 & 5 \\
    $g_1$ & 1 & 0 & 3 & 2 & 5 & 4 \\
    $g_2$ & 2 & 4 & 0 & 5 & 1 & 3 \\
    $g_3$ & 3 & 5 & 1 & 4 & 0 & 2 \\
    $g_4$ & 4 & 2 & 5 & 0 & 3 & 1 \\
    $g_5$ & 5 & 3 & 4 & 1 & 2 & 0 \\
  \end{tabular}
\end{center}
We denote $G(i,j)=g_i g_j$, e.g.
\begin{align}
  G(1,2) &= g_1 g_2 = g_3 := 3, \notag \\
  G(2,1) &= g_2 g_1 = g_4 := 4.
\end{align}
Now the dimension and rank of matrix $\delta_{G(i,j),k}$ is $36\times6$ and $6$, respectively. The null space of $\delta_{G(i,j),k}^{\mathrm{T}}$ is thus spanned by 30 vectors $v^{(p)}$ such that
\begin{equation}
  v^{(p)} = \begin{pmatrix}
    \tilde{v}^{(p)} \\ \hat{v}^{(p)}
  \end{pmatrix}, \quad p \in \{0, 1, \dots, 29\},
\end{equation}
where $\tilde{v}^{(p)}$ are length-6 vectors:
\def\V#1{\tilde{v}^{(#1)}}
\begin{align}
  \V{1} = \V{8}  = \V{16} = \V{21} = \V{29} &= \bigl( 1, 0, 0, 0, 0, 0 \bigr)^{\mathrm{T}}, \notag \\
  \V{0} = \V{10} = \V{14} = \V{23} = \V{27} &= \bigl( 0, 1, 0, 0, 0, 0 \bigr)^{\mathrm{T}}, \notag \\
  \V{3} = \V{6}  = \V{17} = \V{19} = \V{28} &= \bigl( 0, 0, 1, 0, 0, 0 \bigr)^{\mathrm{T}}, \notag \\
  \V{2} = \V{11} = \V{12} = \V{22} = \V{25} &= \bigl( 0, 0, 0, 1, 0, 0 \bigr)^{\mathrm{T}}, \notag \\
  \V{5} = \V{7}  = \V{15} = \V{18} = \V{26} &= \bigl( 0, 0, 0, 0, 1, 0 \bigr)^{\mathrm{T}}, \notag \\
  \V{4} = \V{9}  = \V{13} = \V{20} = \V{24} &= \bigl( 0, 0, 0, 0, 0, 1 \bigr)^{\mathrm{T}}
\end{align}
and $\hat{v}^{(p)}$ are length-30 vectors:
\begin{equation}
  \hat{v}^{(p)}_q = \delta_{pq}, \quad p,q \in \{0, 1, \dots, 29\}.
\end{equation}
The general solution is then given by the linear combination of these $v^{(p)}$ plus the specific part $A^{(\mu)}_{(ij),(0j')}=(\sigma_\mu)_{G(i,j),j'}$ as in Equation~\eqref{eq:1+1d-specific-solution}.

For the solution leading to simple forms in the reconstruction, the coefficients in Equation~\eqref{eq:1+1d-simple-form-augmented-constraint} form a $216\times72$ matrix $\tilde{M}$, whose rank is 66. Hence there are 6 solutions for $A=\tilde{A}_L\otimes\I$, which are given by
\begin{align}
  \tilde{A}^{(0)}_L = B^{(0)}_L &= \begin{pmatrix}
    1 & 0 & 0 & 0 & 0 & 0 \\
    0 & 1 & 0 & 0 & 0 & 0 \\
    0 & 0 & 1 & 0 & 0 & 0 \\
    0 & 0 & 0 & 1 & 0 & 0 \\
    0 & 0 & 0 & 0 & 1 & 0 \\
    0 & 0 & 0 & 0 & 0 & 1 \\
  \end{pmatrix}, \notag \\
  \tilde{A}^{(1)}_L = B^{(1)}_L &= \begin{pmatrix}
    0 & 1 & 1 & 1 & 1 & 1 \\
    1 & 0 & 1 & 1 & 1 & 1 \\
    1 & 1 & 0 & 1 & 1 & 1 \\
    1 & 1 & 1 & 0 & 1 & 1 \\
    1 & 1 & 1 & 1 & 0 & 1 \\
    1 & 1 & 1 & 1 & 1 & 0 \\
  \end{pmatrix}, \notag \\
  \tilde{A}^{(2)}_L = B^{(2)}_L &= \begin{pmatrix}
    0 & -1 & 1 & 1 & 1 & 1 \\
    -1 & 0 & 1 & 1 & 1 & 1 \\
    1 & 1 & 0 & -1 & 1 & 1 \\
    1 & 1 & -1 & 0 & 1 & 1 \\
    1 & 1 & 1 & 1 & 0 & -1 \\
    1 & 1 & 1 & 1 & -1 & 0 \\
  \end{pmatrix}, \notag \\
  \tilde{A}^{(3)}_L = B^{(3)}_L &= \begin{pmatrix}
    0 & 0 & 1 & 0 & 0 & 0 \\
    0 & 0 & 0 & 0 & 1 & 0 \\
    1 & 0 & 0 & 0 & 0 & 0 \\
    0 & 0 & 0 & 0 & 0 & 1 \\
    0 & 1 & 0 & 0 & 0 & 0 \\
    0 & 0 & 0 & 1 & 0 & 0 \\
  \end{pmatrix}, \notag \\
  \tilde{A}^{(4)}_L = B^{(4)}_L &= \begin{pmatrix}
    0 & 0 & \frac12 & -\omega & \omega & -1 \\[0.5ex]
    0 & 0 & \omega & -1 & \frac12 & -\omega \\[0.5ex]
    \frac12 & -\omega & 0 & 0 & -1 & \omega \\[0.5ex]
    \omega & -1 & 0 & 0 & -\omega & \frac12 \\[0.5ex]
    -\omega & \frac12 & -1 & \omega & 0 & 0 \\[0.5ex]
    -1 & \omega & -\omega & \frac12 & 0 & 0 \\
  \end{pmatrix}, \notag \\
  \tilde{A}^{(5)}_L = B^{(5)}_L &= \begin{pmatrix}
    0 & 0 & \frac52 & \eta & \bar{\eta} & 1 \\[0.5ex]
    0 & 0 & \bar{\eta} & 1 & \frac52 & \eta \\[0.5ex]
    \frac52 & \eta & 0 & 0 & 1 & \bar{\eta} \\[0.5ex]
    \bar{\eta} & 1 & 0 & 0 & \eta & \frac52 \\[0.5ex]
    \eta & \frac52 & 1 & \bar{\eta} & 0 & 0 \\[0.5ex]
    1 & \bar{\eta} & \eta & \frac52 & 0 & 0 \\
  \end{pmatrix},
\end{align}
where $\omega=\ee^{\pi\ii/3}$ and $\eta=-\frac{1+3\sqrt3\ii}{2}$. Since $S_3$ is non-abelian, we need to separately solve for $A=\I\otimes\tilde{A}_R$, where the solutions are
\begin{align}
  \tilde{A}^{(0)}_R = B^{(0)}_R &= \begin{pmatrix}
    1 & 0 & 0 & 0 & 0 & 0 \\
    0 & 1 & 0 & 0 & 0 & 0 \\
    0 & 0 & 1 & 0 & 0 & 0 \\
    0 & 0 & 0 & 1 & 0 & 0 \\
    0 & 0 & 0 & 0 & 1 & 0 \\
    0 & 0 & 0 & 0 & 0 & 1 \\
  \end{pmatrix}, \notag \\
  \tilde{A}^{(1)}_R = B^{(1)}_R &= \begin{pmatrix}
    0 & 1 & 1 & 1 & 1 & 1 \\
    1 & 0 & 1 & 1 & 1 & 1 \\
    1 & 1 & 0 & 1 & 1 & 1 \\
    1 & 1 & 1 & 0 & 1 & 1 \\
    1 & 1 & 1 & 1 & 0 & 1 \\
    1 & 1 & 1 & 1 & 1 & 0 \\
  \end{pmatrix}, \notag \\
  \tilde{A}^{(2)}_R = B^{(2)}_R &= \begin{pmatrix}
    0 & 1 & 1 & 1 & 1 & -2 \\
    1 & 0 & 1 & 1 & -2 & 1 \\
    1 & 1 & 0 & -2 & 1 & 1 \\
    1 & 1 & -2 & 0 & 1 & 1 \\
    1 & -2 & 1 & 1 & 0 & 1 \\
    -2 & 1 & 1 & 1 & 1 & 0 \\
  \end{pmatrix}, \notag \\
  \tilde{A}^{(3)}_R = B^{(3)}_R &= \begin{pmatrix}
    0 & 0 & 0 & 1 & 0 & 0 \\
    0 & 0 & 1 & 0 & 0 & 0 \\
    0 & 0 & 0 & 0 & 0 & 1 \\
    0 & 0 & 0 & 0 & 1 & 0 \\
    1 & 0 & 0 & 0 & 0 & 0 \\
    0 & 1 & 0 & 0 & 0 & 0 \\
  \end{pmatrix}, \notag \\
  \tilde{A}^{(4)}_R = B^{(4)}_R &= \begin{pmatrix}
    0 & -\omega & \omega^2 & 1 & 1 & -\frac12 \\[0.5ex]
    -\omega & 0 & 1 & \omega^2 & -\frac12 & 1 \\[0.5ex]
    \omega^2 & 1 & 0 & -\frac12 & -\omega & 1 \\[0.5ex]
    1 & \omega^2 & -\frac12 & 0 & 1 & -\omega \\[0.5ex]
    1 & -\frac12 & -\omega & 1 & 0 & \omega^2 \\[0.5ex]
    -\frac12 & 1 & 1 & -\omega & \omega^2 & 0 \\
  \end{pmatrix}, \notag \\
  \tilde{A}^{(5)}_R = B^{(5)}_R &= \begin{pmatrix}
    0 & \xi & \bar{\xi} & 1 & 1 & 1 \\
    \xi & 0 & 1 & \bar{\xi} & 1 & 1 \\
    \bar{\xi} & 1 & 0 & 1 & \xi & 1 \\
    1 & \bar{\xi} & 1 & 0 & 1 & \xi \\
    1 & 1 & \xi & 1 & 0 & \bar{\xi} \\
    1 & 1 & 1 & \xi & \bar{\xi} & 0 \\
  \end{pmatrix},
\end{align}
where $\xi=-2+\sqrt3\ii$. It can be seen that the number of solutions is equal to the number of group elements as in $\mathbb{Z}_n$ case. In addition, all the $\tilde{A}$ falls in the set of $B$, so the operator pushing procedure can be iterated to high levels as well.

We also note that two solutions in each part coincide, i.e.\ $\tilde{A}^{(0)}_L=\tilde{A}^{(0)}_R$ and $\tilde{A}^{(1)}_L=\tilde{A}^{(1)}_R$. It indicates the abelian $\mathbb{Z}_2$ subgroup of $S_3$.

\section{Operator pushing in 2+1d}

\begin{figure}[htb]
  \centering
  \begin{tikzpicture}
    \def\PathI{(0,1.6) -- (3.2,3.2)}
    \def\PathII{
      (0,1.6) -- (2,0) -- (5,1.6) -- (2.7,5.4) -- cycle
      (2.7,5.4) -- (2,0)
      (2.7,5.4) -- (3.2,3.2) -- (5,1.6)
      (2,0) -- (3.2,3.2)
    }
    \begin{scope}
      \draw [thick, dashed, MaterialGrey] \PathI;
      \draw [thick, MaterialGrey] \PathII;
      \draw (0.8,4  ) node {$i$}
            (4.5,3.8) node {$j$}
            (1.7,1.3) node {$m$}
            (5.5,5.4) node {$n$}
            (3  ,1.4) node {$a$}
            (0.6,0.3) node {$b$}
            (1.6,2.9) node {$c$}
            (4.0,0.3) node {$d$}
            (3.7,2.1) node {$e$};
      \draw [MaterialGrey, -Stealth] (3.1,4.2) to [out=30, in=180] (5,5.4);
      \draw (2.5,-1.5) node {(a)};
    \end{scope}
    \begin{scope}[xshift=10em]
      \draw [thick, dashed, MaterialGrey, ->-=0.6] \PathI;
      \draw [thick, MaterialGrey, ->-=0.055, ->-=0.3, ->-=0.61, ->-=0.96] \PathII;
      \draw [thick, MaterialGrey, ->-=0.2, ->-=0.51, ->-=0.68, ->-=0.9]
        (0,1.6) -- (2.7,5.4) (2.7,5.4) -- (3.2,3.2) (5,1.6) -- (3.2,3.2) (5,1.6) -- (2,0);
      \draw (2.5,-1.5) node {(b)};
    \end{scope}
  \end{tikzpicture}
  \caption{(a) A tetrahedron in the 2+1d tensor network. (b) Arrows on each edge indicate the direction of fusions. If all the objects are self-dual, then these arrows can be ignored.}
  \label{fig:tetrahedra}
\end{figure}
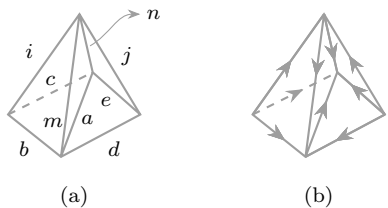

The construction and analysis in the previous section have a natural generalization in one higher dimension. Families of 3d (or 2+1d) holographic tensor networks can be constructed from Levin--Wen string net models. The holographic tensor networks constructed from Levin-Wen models have been discussed in~\cite{chen2022exact}. The unit tensor constituting the tensor network is represented as a tetrahedron drawn in \autoref{fig:tetrahedra}.

As a coarse graining map, it takes edges $i,j,m,n$ to $a$ (with $b,c,d,e$ as spectators untouched). Therefore the problem of operator reconstruction can be understood as reconstructing the bulk operator $B$ acting on leg $a$ by operators $A$ acting on legs $i,j,m,n$. We would also like to find the set of operators that are generalized free operators.

Each tetrahedron has 6 edges which are labeled by the objects in a fusion category characterizing the topological field theory. For a given labeled tetrahedron, it is assigned the value of an $F$-symbol depending on the labels on the 6 edges. It is non-vanishing if there exists a fusion channel for the three edges of every triangle to fuse to the trivial identity object. Because of this constraint, it is more convenient to think of each allowed configuration of a triangle as our fundamental degrees of freedom (in the bulk) and to consider operators acting on the triangle that transform them between allowed configurations. The problem of operator pushing across the unit tensor can be formulated as pushing the bulk operator action on the triangle basis $\triangle_{abc}$ or $\triangle_{ade}$.

Since the above diagram is symmetric, we only need to consider one tetrahedron such as $abcinm$ and the other can be simply obtained by flipping all legs. Here we use the following convention: lowercase letters such as $i,j\in\{0,1,\dots,n-1\}$ are reserved for edge labels, where $n$ is the number of objects in the fusion category; uppercase $I, J\in \{0,1,\dots,N-1\}$ are triangle (or face, or fusion channel) labels, and $N$ is the number of admissible configurations on a triangle. The tetrahedron can then be labeled as
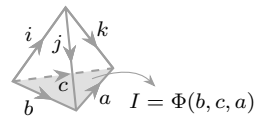
\[
  \begin{tikzpicture}
    \fill [MaterialGrey300]
          (0,0) -- (3.8,0.5) -- (2.4,-1.1) -- cycle;
    \draw [thick, dashed, MaterialGrey, ->-=0.58] (0,0) -- (3.8,0.5);
    \draw [thick, MaterialGrey, ->-=0.11, ->-=0.27, ->-=0.88]
          (0,0) -- (2.4,-1.1) -- (3.8,0.5) -- (2,2.8) -- cycle
          (2,2.8) -- (2.4,-1.1);
    \draw [thick, MaterialGrey, ->-=0.32, ->-=0.85]
          (0,0) -- (2,2.8) -- (3.8,0.5);
    \draw (0.6, 1.8) node {$i$}
          (1.7, 1.4) node {$j$}
          (3.4, 2.0) node {$k$}
          (3.5,-0.7) node {$a$}
          (0.6,-1.0) node {$b$}
          (1.9,-0.2) node {$c$}
          (6.8,-0.8) node {$I=\Phi(b,c,a)$};
    \draw [MaterialGrey, -Stealth] (3,0.1) to [out=20] (5.4,-0.2);
  \end{tikzpicture}
\]
where $I=\Phi(b,c,a)$ is the triangle label, and function $\Phi$ can be determined by fusion rules.

The constraint equation in 2+1d is similar to the 1+1d case Equation~\eqref{eq:1+1d-constraint} and \eqref{eq:1+1d-constraint-in-sigma}:
\begin{align}
  &\mathrel{\phantom{=}}
     A_{(ijk), (i'j'k')} M_{(i'j'k'), I} \notag \\
  &= M_{(ijk), I'} B_{I'I}
   = M_{(ijk), I'} (\sigma_\mu)_{I'I},
\end{align}
where $\mu\in\{0,1,\dots,N^2-1\}$ labels the generalized Pauli matrices in the triangle basis. Now $M$ is determined by
\begin{equation}
  M_{(ijk), I} = M_{(ijk), \Phi(b,c,a)} = \sqrt{d_j d_k d_b d_c} \, \bigl[ F^{jkb}_c \bigr]_{ia},
\end{equation}
where $[F^{jkb}_c]_{ia}$ is the $F$-symbol and $d_i$ is quantum dimension of object $i$ in the fusion category.

Similarly, to check if $A$ can be written in a ``simple'' form, we look for $A$ of the form say:
\begin{equation}
  A_{(ijk), (i'j'k')} = \tilde{A}_{ii'} \delta_{jj'} \delta_{kk'},
\end{equation}
such that
\begin{multline}
  \tilde{A}_{ii'} M_{(i'jk), I} = M_{(ijk), I'} (\sigma_\mu)_{I'I}, \\
  \forall j, k \in \{ 0, 1, \dots, n-1 \}.
\end{multline}
The condition for simple form reconstruction is thus
\begin{align}
  &\mathrel{\phantom{=}}
     \rank\bigl[ (M_{(i'jk), I})^{\mathrm{T}} | (M_{(ijk), I'} (\sigma_\mu)_{I'I})^{\mathrm{T}}_s \bigr] \notag \\
  &= \rank\bigl( M_{(i'jk), I} \bigr).
\end{align}
We can also change it to
\begin{equation}
  \tilde{A}_{ii'} M_{(i'jk), I} - \sum_\mu \alpha_\mu M_{(ijk), I'} (\sigma_\mu)_{I'I} = 0
  \label{eq:2+1d-simple-form-augmented-constraint}
\end{equation}
to explicitly solve $B$. Taking indices $i$, $j$ or $i$, $k$ to be identical mappings will lead to other simple form solutions as well.

\subsection{\texorpdfstring{$\mathbb{Z}_n$}{ℤₙ} case}

We again look into the simple example where the fusion category is the abelian group $\mathbb{Z}_n$. The objects that label the edges of the tetrahedron are again taken from group elements of $\mathbb{Z}_n$. Recall that in each admissible triangle two of the edges must fuse to the third, and fusion here again means group product of $\mathbb{Z}_n$, which has been introduced in Equation~\eqref{eq:zn-fusion-rules}. The number of admissible triangles is thus given by $N=n^2$, i.e.\ the triangles labels lie in $\{0,1,\dots,n^2-1\}$. Taking $\mathbb{Z}_2$ as an example, the admissible triangles (it is also customary to use the dual graph of the triangle to highlight the fusion relation between the edges, which we adopt below) are listed in the following. We also assign a label to each admissible triangle, from 0 to 3.
\begin{equation}
  \begin{gathered}
    \Fusion000 \to 0, \quad
    \Fusion011 \to 1, \\
    \Fusion101 \to 2, \quad
    \Fusion110 \to 3.
  \end{gathered}
  \label{eq:fusion-z2}
\end{equation}
The above triangle labeling rule for generic $n$ can be stated as:
\begin{equation}
  I = nb + c = n [(i+j)\bmod n] + [(i+k)\bmod n],
\end{equation}
where $i,j$ are two of the edges of the triangle [say the two top edges when adopting notation as in Equation~\eqref{eq:fusion-z2}]. Consider the class of trivial $\mathbb{Z}_n$ Dijkgraaf--Witten models, where all $F$-symbols equal unity when fusion constraints at each triangle are satisfied~\footnote{For generic $\mathbb{Z}_n$ group where the group elements are generically not self-dual, one has to put in an orientation on each edge [see \autoref{fig:tetrahedra}(b)]. This can be done by numbering the vertices, and attaching an arrow to each edge pointing from the vertex with a smaller number to the other with a larger label. The result is independent of such labeling.}. We have
\begin{equation}
  M_{(ijk), I} = \delta_{n[(i+j)\bmod n]+[(i+k)\bmod n], I}.
\end{equation}
So the constraint equation becomes
\begin{align}
  &\mathrel{\phantom{=}}
     A_{(ijk), (i'j'k')} \delta_{n[(i'+j')\bmod n]+[(i'+k')\bmod n], I} \notag \\
  &= (\sigma_\mu)_{n[(i+j)\bmod n]+[(i+k)\bmod n], I}.
\end{align}
Now $M^{\mathrm{T}}$ is a rank-$n^2$ matrix, whose null space is spanned by $n^3-n^2$ vectors $v^{(p)}$ such that:
\begin{align}
  v^{(p)}_q &= \delta_{n [(- \lfloor p/n\rfloor - \lfloor p/n^2 \rfloor - 2) \bmod n] + [(- \lfloor p/n^2 \rfloor - 2) \bmod n], q} \notag \\
  &\quad {} - \delta_{n^3-p-1, q},
\end{align}
where $p\in\{0,1,\dots,n^3-n^2-1\}$, $q\in\{0,1,\dots,n^3-1\}$ and the specific solution part is
\begin{equation}
  A^{(\mu)}_{(ijk), (0j'k')} = (\sigma_\mu)_{n[(i+j)\bmod n]+[(i+k)\bmod n], nj'+k'}.
\end{equation}

For the simple operator case, we check Equation~\eqref{eq:2+1d-simple-form-augmented-constraint}, whose coefficients form an $n^5\times(n^4+n^2)$ matrix with rank $n^4+n^2-n$, so its null space is spanned by $n$ vectors. For general $n$, it's difficult to solve Equation~\eqref{eq:2+1d-simple-form-augmented-constraint} and we only consider small $n$ here. For $n=2$, we have
\begin{equation}
  \begin{aligned}
    B^{(0)} &= \begin{pmatrix}
      1 & 0 & 0 & 0 \\
      0 & 1 & 0 & 0 \\
      0 & 0 & 1 & 0 \\
      0 & 0 & 0 & 1 \\
    \end{pmatrix}
    = \sigma^{(2)}_0 \otimes \sigma^{(2)}_0, \\
    B^{(1)} &= \begin{pmatrix}
      0 & 0 & 0 & 1 \\
      0 & 0 & 1 & 0 \\
      0 & 1 & 0 & 0 \\
      1 & 0 & 0 & 0 \\
    \end{pmatrix}
    = \sigma^{(2)}_1 \otimes \sigma^{(2)}_1
  \end{aligned}
  \label{eq:2+1d-z2-solution-b}
\end{equation}
and
\begin{equation}
  \tilde{A}^{(0)} = \sigma^{(2)}_0, \quad
  \tilde{A}^{(1)} = \sigma^{(2)}_1.
  \label{eq:2+1d-z2-solution-a}
\end{equation}
For $n=3$, the solutions are
\begin{equation}
  \begin{aligned}
    B^{(0)} &= \sigma^{(3)}_0 \otimes \sigma^{(3)}_0, \\
    B^{(1)} &= \sigma^{(3)}_1 \otimes \sigma^{(3)}_1, \\
    B^{(2)} &= \sigma^{(3)}_2 \otimes \sigma^{(3)}_2
  \end{aligned}
  \label{eq:2+1d-z3-solution-b}
\end{equation}
and
\begin{equation}
  \tilde{A}^{(0)} = \sigma^{(3)}_0, \quad
  \tilde{A}^{(1)} = \sigma^{(3)}_1, \quad
  \tilde{A}^{(2)} = \sigma^{(3)}_2.
  \label{eq:2+1d-z3-solution-a}
\end{equation}
Here, we use the superscript of $\sigma$ to denote its size for clarity.

Other simple form solutions can be calculated in the same way. When $A_{(ijk),(i'j'k')}=\tilde{A}_{jj'}\delta_{ii'}\delta_{kk'}$, we have
\begin{equation}
  B^{(0)} = \sigma^{(2)}_0 \otimes \sigma^{(2)}_0, \quad
  B^{(1)} = \sigma^{(2)}_1 \otimes \sigma^{(2)}_0
\end{equation}
for $\mathbb{Z}_2$ and
\begin{equation}
  \begin{aligned}
    B^{(0)} &= \sigma^{(3)}_0 \otimes \sigma^{(3)}_0, \\
    B^{(1)} &= \sigma^{(3)}_1 \otimes \sigma^{(3)}_0, \\
    B^{(2)} &= \sigma^{(3)}_2 \otimes \sigma^{(3)}_0
  \end{aligned}
\end{equation}
for $\mathbb{Z}_3$; when $A_{(ijk),(i'j'k')}=\tilde{A}_{kk'}\delta_{ii'}\delta_{jj'}$, we have
\begin{equation}
  B^{(0)} = \sigma^{(2)}_0 \otimes \sigma^{(2)}_0, \quad
  B^{(1)} = \sigma^{(2)}_0 \otimes \sigma^{(2)}_1
\end{equation}
for $\mathbb{Z}_2$ and
\begin{equation}
  \begin{aligned}
    B^{(0)} &= \sigma^{(3)}_0 \otimes \sigma^{(3)}_0, \\
    B^{(1)} &= \sigma^{(3)}_0 \otimes \sigma^{(3)}_1, \\
    B^{(2)} &= \sigma^{(3)}_0 \otimes \sigma^{(3)}_2
  \end{aligned}
\end{equation}
for $\mathbb{Z}_3$. The corresponding $\tilde{A}$ are the same as in Equation~\eqref{eq:2+1d-z2-solution-a} and \eqref{eq:2+1d-z3-solution-a}.

In \autoref{fig:tetrahedra}, $B$ operators in the two tetrahedra will act on triangles $\triangle_{abc}$ and $\triangle_{ade}$. In the above equations, we see $B$ can be decomposed into small $\sigma$ that act on the edges. For the next iteration of operator pushing, where the bulk is now given by triangles $\triangle_{bim}$, $\triangle_{cin}$, $\triangle_{djm}$ and $\triangle_{ejn}$, it can be seen that the $A$ operators acting on edge $i$ and $j$, as well as the decomposed $B$ operators on edge $b$, $c$, $d$, $e$, will altogether give the new $B$ operator for these four triangles. Therefore, the operator pushing procedure can be reiterated at the next level.

\subsection{Fibonacci model}
Another important example of topological order in 2+1 dimension is the Fibonacci model. As a fusion category, there are two objects $1,\tau$, and they satisfy the following fusion rules:
\begin{equation}
  1 \otimes 1 = 1, \quad
  1 \otimes \tau = \tau \otimes 1 = \tau, \quad
  \tau \otimes \tau = 1 \oplus \tau
\end{equation}
so there are 5 admissible triangles:
\begin{equation}
  \begin{gathered}
    \Fusion000 \to 0, \quad
    \Fusion011 \to 1, \quad
    \Fusion101 \to 2, \\
    \Fusion110 \to 3, \quad
    \Fusion111 \to 4,
  \end{gathered}
\end{equation}
and the non-vanishing $F$ symbols are
\begin{equation}
  F^{\tau\tau\tau}_\tau = \begin{pmatrix}
    \phi^{-1} & \phi^{-1/2} \\
    \phi^{-1/2} & -\phi^{-1}
  \end{pmatrix}
\end{equation}
where $\phi = (1+\sqrt5)/2$ denotes the golden ratio. Then
\begin{equation}
  M_{(ijk), I} = \begin{pmatrix}
    1 & 0 & 0 & 0 & 0 \\
    0 & 0 & \phi & 0 & 0 \\
    0 & 0 & 0 & \phi & 0 \\
    0 & \phi & 0 & 0 & \phi^{3/2} \\
    0 & \phi & 0 & 0 & 0 \\
    0 & 0 & 0 & \phi & \phi^{3/2} \\
    0 & 0 & \phi & 0 & \phi^{3/2} \\
    \phi & \phi^{3/2} & \phi^{3/2} & \phi^{3/2} & -\phi
  \end{pmatrix}.
\end{equation}
We can see $\rank(M)=5$ so the null space of $M^{\mathrm{T}}$ is spanned by
\begin{equation}
  \{ v^{(p)} \} = \Biggl\{
    \begin{pmatrix} \phi^{1/2} \\ 1 \\ 1 \\ -\phi^{-1} \\ \phi \\ 0 \\ 0 \\ -\phi^{1/2} \end{pmatrix}, \,
    \begin{pmatrix} 0 \\ 1 \\ 0 \\ 1 \\ -1 \\  0 \\ -1 \\ 0 \end{pmatrix}, \,
    \begin{pmatrix} 0 \\ 0 \\ 1 \\ 1 \\ -1 \\ -1 \\  0 \\ 0 \end{pmatrix}
  \Biggr\},
  \label{eq:2+1d-fib-null-space}
\end{equation}
The specific solution part for each $\sigma_\mu$ can thus be calculated, see Appendix~\ref{sec:2+1d-fib-solution}.

For the simple form, we can see the coefficients matrix of Equation~\eqref{eq:2+1d-simple-form-augmented-constraint} is $(2^3\times5)\times(2^2+5^2)=40\times29$, but it has rank 28, so there is only one solution. However, since the trivial solution ($\tilde{A}$ and $B$ are both identity operators) always exists, there doesn't exist non-trivial generalized free field corresponding to
\begin{equation}
  A = \tilde{A} \otimes \I \otimes \I \enspace \text{or} \enspace
      \I \otimes \tilde{A} \otimes \I \enspace \text{or} \enspace
      \I \otimes \I \otimes \tilde{A}
\end{equation}
in the Fibonacci model.

\section{Conclusions}

In this paper, we explore the emergence of generalized free fields in classes of holographic tensor networks constructed from topological field theories. We considered both 1+1 dimensional and 2+1 dimension networks and demonstrated for example in networks following from abelian Dijkgraaf--Witten theories in the trivial cohomological class, the number of generalized free fields scale with the rank of the group. Interestingly, for the simple case of a Fibonacci model, its RG network admits no generalized free field. Of course, to recover a holographic network that resembles a semi-classical bulk theory, it is expected that the spectrum has a large gap with only a sparse number of free fields as the degrees of freedom (i.e.\ central charge) approaches infinity. While we do not expect such a simple model to admit generalized free fields, it is interesting to find out under what circumstances they would admit some. 

These models are not expected to recover a semi-classical bulk theory, although they are examples where the dual CFT can be constructed explicitly. It would be interesting to study more generic TQFT with other large ``rank'' limit to look for bulk networks with semi-classical approximations.

\begin{acknowledgments}
  LYH acknowledges the support of NSFC (Grant No.\ 11922502, 11875111). We also thank Xinyang Yu, Yan\-yan Chen, and Tian Yuan for useful discussions and comments.
\end{acknowledgments}

\bibliography{main}

\onecolumngrid

\appendix

\section{Solution for 2+1d Fibonacci model}
\label{sec:2+1d-fib-solution}

The general solution for the 2+1d Fibonacci model is given by the linear combination of vectors $v^{(p)}$ in Equation~\eqref{eq:2+1d-fib-null-space} plus the specific solution part $A^{(\mu)}$ for each $\sigma_\mu$. $A^{(\mu)}$ is an $8\times8$ matrix whose last 3 columns are zero, i.e.
\begin{equation}
  A^{(\mu)} = \begin{pmatrix} \tilde{A}^{(\mu)} & \textbf{0} & \textbf{0} & \textbf{0} \end{pmatrix},
\end{equation}
and the $8\times5$ matrices $\tilde{A}^{(\mu)}$ are given by
\begin{align}
  \tilde{A}^{(0)} &= \begin{pmatrix}
    1 & 0 & 0 & 0 & 0 \\
    0 & 1 & 0 & 0 & 0 \\
    0 & 0 & 1 & 0 & 0 \\
    0 & 0 & 0 & 1 & 0 \\
    0 & 0 & 0 & 0 & 1 \\
    0 & 0 & 1 & 1 & -1 \\
    0 & 1 & 0 & 1 & -1 \\
    \phi & \sqrt{\phi} & \sqrt{\phi} & -\frac{1}{\sqrt{\phi}} & \frac{\phi+1}{\sqrt{\phi}} \\
  \end{pmatrix}, &
  \tilde{A}^{(1)} &= \begin{pmatrix}
    0 & 0 & 0 & 0 & \frac{1}{\phi} \\
    0 & 0 & 1 & 0 & 0 \\
    0 & 0 & 0 & \frac{1}{\sqrt{\phi}} & -\frac{1}{\sqrt{\phi}} \\
    \phi^{3/2} & 1 & 0 & 0 & 0 \\
    0 & 1 & 0 & 0 & 0 \\
    \phi^{3/2} & 0 & 0 & \frac{1}{\sqrt{\phi}} & -\frac{1}{\sqrt{\phi}} \\
    \phi^{3/2} & 0 & 1 & 0 & 0 \\
    -\phi & \sqrt{\phi} & \sqrt{\phi} & 1 & 0 \\
  \end{pmatrix}, \notag \\[1ex]
  \tilde{A}^{(2)} &= \begin{pmatrix}
    0 & \frac{1}{\phi} & 0 & 0 & 0 \\
    0 & 0 & 0 & \frac{1}{\sqrt{\phi}} & -\frac{1}{\sqrt{\phi}} \\
    \phi & 0 & 0 & 0 & 0 \\
    0 & 0 & 1 & 0 & \sqrt{\phi} \\
    0 & 0 & 1 & 0 & 0 \\
    \phi & 0 & 0 & 0 & \sqrt{\phi} \\
    0 & 0 & 0 & \frac{1}{\sqrt{\phi}} & \frac{\phi-1}{\sqrt{\phi}} \\
    \phi^{3/2} & 1 & \sqrt{\phi} & 1 & -2 \\
  \end{pmatrix}, &
  \tilde{A}^{(3)} &= \begin{pmatrix}
    0 & 0 & \frac{1}{\phi} & 0 & 0 \\
    \phi & 0 & 0 & 0 & 0 \\
    0 & 0 & 0 & 0 & 1 \\
    0 & \sqrt{\phi} & 0 & \frac{1}{\sqrt{\phi}} & -\frac{1}{\sqrt{\phi}} \\
    0 & 0 & 0 & \frac{1}{\sqrt{\phi}} & -\frac{1}{\sqrt{\phi}} \\
    0 & \sqrt{\phi} & 0 & 0 & 1 \\
    \phi & \sqrt{\phi} & 0 & 0 & 0 \\
    \phi^{3/2} & -1 & 1 & 1 & \sqrt{\phi}-1 \\
  \end{pmatrix}, \notag \\[1ex]
  \tilde{A}^{(4)} &= \begin{pmatrix}
    0 & 0 & 0 & \frac{1}{\phi^{3/2}} & -\frac{1}{\phi^{3/2}} \\
    0 & 0 & 0 & 0 & 1 \\
    0 & 1 & 0 & 0 & 0 \\
    \phi & 0 & \sqrt{\phi} & 0 & 0 \\
    \phi & 0 & 0 & 0 & 0 \\
    0 & 1 & \sqrt{\phi} & 0 & 0 \\
    0 & 0 & \sqrt{\phi} & 0 & 1 \\
    \phi^{3/2} & \sqrt{\phi} & -1 & \frac{1}{\sqrt{\phi}} & \frac{\phi-1}{\sqrt{\phi}} \\
  \end{pmatrix}, &
  \tilde{A}^{(5)} &= \begin{pmatrix}
    1 & 0 & 0 & 0 & 0 \\
    0 & \omega^4 & 0 & 0 & 0 \\
    0 & 0 & -\omega & 0 & 0 \\
    0 & 0 & 0 & -\omega^3 & \omega^3+\omega^2 \\
    0 & 0 & 0 & 0 & \omega^2 \\
    0 & 0 & -\omega & -\omega^3 & \omega^3 \\
    0 & \omega^4 & 0 & -\omega^3 & \omega^3 \\
    \phi & \omega^4 \sqrt{\phi} & -\omega\sqrt{\phi} & \frac{\omega^3}{\sqrt{\phi}} & \frac{\omega^2 \phi-\omega^3}{\sqrt{\phi}} \\
  \end{pmatrix}, \notag \\[1ex]
  \tilde{A}^{(6)} &= \begin{pmatrix}
    0 & 0 & 0 & 0 & \frac{\omega^2}{\phi} \\
    0 & 0 & -\omega & 0 & 0 \\
    0 & 0 & 0 & -\frac{\omega^3}{\sqrt{\phi}} & \frac{\omega^3}{\sqrt{\phi}} \\
    \phi^{3/2} & \omega^4 & 0 & 0 & 0 \\
    0 & \omega^4 & 0 & 0 & 0 \\
    \phi^{3/2} & 0 & 0 & -\frac{\omega^3}{\sqrt{\phi}} & \frac{\omega^3}{\sqrt{\phi}} \\
    \phi^{3/2} & 0 & -\omega & 0 & 0 \\
    -\phi & \omega^4 \sqrt{\phi} & -\omega\sqrt{\phi} & -\omega^3 & \omega^3+\omega^2 \\
  \end{pmatrix}, &
  \tilde{A}^{(7)} &= \begin{pmatrix}
    0 & \frac{\omega^4}{\phi} & 0 & 0 & 0 \\
    0 & 0 & 0 & -\frac{\omega^3}{\sqrt{\phi}} & \frac{\omega^3}{\sqrt{\phi}} \\
    \phi & 0 & 0 & 0 & 0 \\
    0 & 0 & -\omega & 0 & \omega^2 \sqrt{\phi} \\
    0 & 0 & -\omega & 0 & 0 \\
    \phi & 0 & 0 & 0 & \omega^2 \sqrt{\phi} \\
    0 & 0 & 0 & -\frac{\omega^3}{\sqrt{\phi}} & \frac{\omega^3+\omega^2 \phi}{\sqrt{\phi}} \\
    \phi^{3/2} & \omega^4 & -\omega\sqrt{\phi} & -\omega^3 & \omega^3-\omega^2 \\
  \end{pmatrix}, \notag \\[1ex]
  \tilde{A}^{(8)} &= \begin{pmatrix}
    0 & 0 & -\frac{\omega}{\phi} & 0 & 0 \\
    \phi & 0 & 0 & 0 & 0 \\
    0 & 0 & 0 & 0 & \omega^2 \\
    0 & \omega^4 \sqrt{\phi} & 0 & -\frac{\omega^3}{\sqrt{\phi}} & \frac{\omega^3}{\sqrt{\phi}} \\
    0 & 0 & 0 & -\frac{\omega^3}{\sqrt{\phi}} & \frac{\omega^3}{\sqrt{\phi}} \\
    0 & \omega^4 \sqrt{\phi} & 0 & 0 & \omega^2 \\
    \phi & \omega^4 \sqrt{\phi} & 0 & 0 & 0 \\
    \phi^{3/2} & -\omega^4 & -\omega & -\omega^3 & \omega^3+\omega^2 \sqrt{\phi} \\
  \end{pmatrix}, &
  \tilde{A}^{(9)} &= \begin{pmatrix}
    0 & 0 & 0 & -\frac{\omega^3}{\phi^{3/2}} & \frac{\omega^3}{\phi^{3/2}} \\
    0 & 0 & 0 & 0 & \omega^2 \\
    0 & \omega^4 & 0 & 0 & 0 \\
    \phi & 0 & -\omega\sqrt{\phi} & 0 & 0 \\
    \phi & 0 & 0 & 0 & 0 \\
    0 & \omega^4 & -\omega\sqrt{\phi} & 0 & 0 \\
    0 & 0 & -\omega\sqrt{\phi} & 0 & \omega^2 \\
    \phi^{3/2} & \omega^4 \sqrt{\phi} & \omega & -\frac{\omega^3}{\sqrt{\phi}} & \frac{\omega^3+\omega^2 \phi}{\sqrt{\phi}} \\
  \end{pmatrix}, \notag \\[1ex]
  \tilde{A}^{(10)} &= \begin{pmatrix}
    1 & 0 & 0 & 0 & 0 \\
    0 & -\omega^3 & 0 & 0 & 0 \\
    0 & 0 & \omega^2 & 0 & 0 \\
    0 & 0 & 0 & -\omega & \omega^4+\omega \\
    0 & 0 & 0 & 0 & \omega^4 \\
    0 & 0 & \omega^2 & -\omega & \omega \\
    0 & -\omega^3 & 0 & -\omega & \omega \\
    \phi & -\omega^3\sqrt{\phi} & \omega^2 \sqrt{\phi} & \frac{\omega}{\sqrt{\phi}} & \frac{\omega^4 \phi-\omega}{\sqrt{\phi}} \\
  \end{pmatrix}, &
  \tilde{A}^{(11)} &= \begin{pmatrix}
    0 & 0 & 0 & 0 & \frac{\omega^4}{\phi} \\
    0 & 0 & \omega^2 & 0 & 0 \\
    0 & 0 & 0 & -\frac{\omega}{\sqrt{\phi}} & \frac{\omega}{\sqrt{\phi}} \\
    \phi^{3/2} & -\omega^3 & 0 & 0 & 0 \\
    0 & -\omega^3 & 0 & 0 & 0 \\
    \phi^{3/2} & 0 & 0 & -\frac{\omega}{\sqrt{\phi}} & \frac{\omega}{\sqrt{\phi}} \\
    \phi^{3/2} & 0 & \omega^2 & 0 & 0 \\
    -\phi & -\omega^3\sqrt{\phi} & \omega^2 \sqrt{\phi} & -\omega & \omega^4+\omega \\
  \end{pmatrix}, \notag \\[1ex]
  \tilde{A}^{(12)} &= \begin{pmatrix}
    0 & -\frac{\omega^3}{\phi} & 0 & 0 & 0 \\
    0 & 0 & 0 & -\frac{\omega}{\sqrt{\phi}} & \frac{\omega}{\sqrt{\phi}} \\
    \phi & 0 & 0 & 0 & 0 \\
    0 & 0 & \omega^2 & 0 & \omega^4 \sqrt{\phi} \\
    0 & 0 & \omega^2 & 0 & 0 \\
    \phi & 0 & 0 & 0 & \omega^4 \sqrt{\phi} \\
    0 & 0 & 0 & -\frac{\omega}{\sqrt{\phi}} & \frac{\omega+\omega^4 \phi}{\sqrt{\phi}} \\
    \phi^{3/2} & -\omega^3 & \omega^2 \sqrt{\phi} & -\omega & \omega-\omega^4 \\
  \end{pmatrix}, &
  \tilde{A}^{(13)} &= \begin{pmatrix}
    0 & 0 & \frac{\omega^2}{\phi} & 0 & 0 \\
    \phi & 0 & 0 & 0 & 0 \\
    0 & 0 & 0 & 0 & \omega^4 \\
    0 & -\omega^3\sqrt{\phi} & 0 & -\frac{\omega}{\sqrt{\phi}} & \frac{\omega}{\sqrt{\phi}} \\
    0 & 0 & 0 & -\frac{\omega}{\sqrt{\phi}} & \frac{\omega}{\sqrt{\phi}} \\
    0 & -\omega^3\sqrt{\phi} & 0 & 0 & \omega^4 \\
    \phi & -\omega^3\sqrt{\phi} & 0 & 0 & 0 \\
    \phi^{3/2} & \omega^3 & \omega^2 & -\omega & \omega+\omega^4 \sqrt{\phi} \\
  \end{pmatrix}, \notag \\[1ex]
  \tilde{A}^{(14)} &= \begin{pmatrix}
    0 & 0 & 0 & -\frac{\omega}{\phi^{3/2}} & \frac{\omega}{\phi^{3/2}} \\
    0 & 0 & 0 & 0 & \omega^4 \\
    0 & -\omega^3 & 0 & 0 & 0 \\
    \phi & 0 & \omega^2 \sqrt{\phi} & 0 & 0 \\
    \phi & 0 & 0 & 0 & 0 \\
    0 & -\omega^3 & \omega^2 \sqrt{\phi} & 0 & 0 \\
    0 & 0 & \omega^2 \sqrt{\phi} & 0 & \omega^4 \\
    \phi^{3/2} & -\omega^3\sqrt{\phi} & -\omega^2 & -\frac{\omega}{\sqrt{\phi}} & \frac{\omega+\omega^4 \phi}{\sqrt{\phi}} \\
  \end{pmatrix}, &
  \tilde{A}^{(15)} &= \begin{pmatrix}
    1 & 0 & 0 & 0 & 0 \\
    0 & \omega^2 & 0 & 0 & 0 \\
    0 & 0 & -\omega^3 & 0 & 0 \\
    0 & 0 & 0 & \omega^4 & -\omega^4-\omega \\
    0 & 0 & 0 & 0 & -\omega \\
    0 & 0 & -\omega^3 & \omega^4 & -\omega^4 \\
    0 & \omega^2 & 0 & \omega^4 & -\omega^4 \\
    \phi & \omega^2 \sqrt{\phi} & -\omega^3\sqrt{\phi} & -\frac{\omega^4}{\sqrt{\phi}} & \frac{\omega^4-\omega \phi}{\sqrt{\phi}} \\
  \end{pmatrix}, \notag \\[1ex]
  \tilde{A}^{(16)} &= \begin{pmatrix}
    0 & 0 & 0 & 0 & -\frac{\omega}{\phi} \\
    0 & 0 & -\omega^3 & 0 & 0 \\
    0 & 0 & 0 & \frac{\omega^4}{\sqrt{\phi}} & -\frac{\omega^4}{\sqrt{\phi}} \\
    \phi^{3/2} & \omega^2 & 0 & 0 & 0 \\
    0 & \omega^2 & 0 & 0 & 0 \\
    \phi^{3/2} & 0 & 0 & \frac{\omega^4}{\sqrt{\phi}} & -\frac{\omega^4}{\sqrt{\phi}} \\
    \phi^{3/2} & 0 & -\omega^3 & 0 & 0 \\
    -\phi & \omega^2 \sqrt{\phi} & -\omega^3\sqrt{\phi} & \omega^4 & -\omega^4-\omega \\
  \end{pmatrix}, &
  \tilde{A}^{(17)} &= \begin{pmatrix}
    0 & \frac{\omega^2}{\phi} & 0 & 0 & 0 \\
    0 & 0 & 0 & \frac{\omega^4}{\sqrt{\phi}} & -\frac{\omega^4}{\sqrt{\phi}} \\
    \phi & 0 & 0 & 0 & 0 \\
    0 & 0 & -\omega^3 & 0 & -\omega\sqrt{\phi} \\
    0 & 0 & -\omega^3 & 0 & 0 \\
    \phi & 0 & 0 & 0 & -\omega\sqrt{\phi} \\
    0 & 0 & 0 & \frac{\omega^4}{\sqrt{\phi}} & \frac{-\omega^4-\omega \phi}{\sqrt{\phi}} \\
    \phi^{3/2} & \omega^2 & -\omega^3\sqrt{\phi} & \omega^4 & \omega-\omega^4 \\
  \end{pmatrix}, \notag \\[1ex]
  \tilde{A}^{(18)} &= \begin{pmatrix}
    0 & 0 & -\frac{\omega^3}{\phi} & 0 & 0 \\
    \phi & 0 & 0 & 0 & 0 \\
    0 & 0 & 0 & 0 & -\omega \\
    0 & \omega^2 \sqrt{\phi} & 0 & \frac{\omega^4}{\sqrt{\phi}} & -\frac{\omega^4}{\sqrt{\phi}} \\
    0 & 0 & 0 & \frac{\omega^4}{\sqrt{\phi}} & -\frac{\omega^4}{\sqrt{\phi}} \\
    0 & \omega^2 \sqrt{\phi} & 0 & 0 & -\omega \\
    \phi & \omega^2 \sqrt{\phi} & 0 & 0 & 0 \\
    \phi^{3/2} & -\omega^2 & -\omega^3 & \omega^4 & -\omega^4-\omega \sqrt{\phi} \\
  \end{pmatrix}, &
  \tilde{A}^{(19)} &= \begin{pmatrix}
    0 & 0 & 0 & \frac{\omega^4}{\phi^{3/2}} & -\frac{\omega^4}{\phi^{3/2}} \\
    0 & 0 & 0 & 0 & -\omega \\
    0 & \omega^2 & 0 & 0 & 0 \\
    \phi & 0 & -\omega^3\sqrt{\phi} & 0 & 0 \\
    \phi & 0 & 0 & 0 & 0 \\
    0 & \omega^2 & -\omega^3\sqrt{\phi} & 0 & 0 \\
    0 & 0 & -\omega^3\sqrt{\phi} & 0 & -\omega \\
    \phi^{3/2} & \omega^2 \sqrt{\phi} & \omega^3 & \frac{\omega^4}{\sqrt{\phi}} & \frac{-\omega^4-\omega \phi}{\sqrt{\phi}} \\
  \end{pmatrix}, \notag \\[1ex]
  \tilde{A}^{(20)} &= \begin{pmatrix}
    1 & 0 & 0 & 0 & 0 \\
    0 & -\omega & 0 & 0 & 0 \\
    0 & 0 & \omega^4 & 0 & 0 \\
    0 & 0 & 0 & \omega^2 & -\omega^3-\omega^2 \\
    0 & 0 & 0 & 0 & -\omega^3 \\
    0 & 0 & \omega^4 & \omega^2 & -\omega^2 \\
    0 & -\omega & 0 & \omega^2 & -\omega^2 \\
    \phi & -\omega\sqrt{\phi} & \omega^4 \sqrt{\phi} & -\frac{\omega^2}{\sqrt{\phi}} & \frac{\omega^2-\omega^3 \phi}{\sqrt{\phi}} \\
  \end{pmatrix}, &
  \tilde{A}^{(21)} &= \begin{pmatrix}
    0 & 0 & 0 & 0 & -\frac{\omega^3}{\phi} \\
    0 & 0 & \omega^4 & 0 & 0 \\
    0 & 0 & 0 & \frac{\omega^2}{\sqrt{\phi}} & -\frac{\omega^2}{\sqrt{\phi}} \\
    \phi^{3/2} & -\omega & 0 & 0 & 0 \\
    0 & -\omega & 0 & 0 & 0 \\
    \phi^{3/2} & 0 & 0 & \frac{\omega^2}{\sqrt{\phi}} & -\frac{\omega^2}{\sqrt{\phi}} \\
    \phi^{3/2} & 0 & \omega^4 & 0 & 0 \\
    -\phi & -\omega\sqrt{\phi} & \omega^4 \sqrt{\phi} & \omega^2 & -\omega^3-\omega^2 \\
  \end{pmatrix}, \notag \\[1ex]
  \tilde{A}^{(22)} &= \begin{pmatrix}
    0 & -\frac{\omega}{\phi} & 0 & 0 & 0 \\
    0 & 0 & 0 & \frac{\omega^2}{\sqrt{\phi}} & -\frac{\omega^2}{\sqrt{\phi}} \\
    \phi & 0 & 0 & 0 & 0 \\
    0 & 0 & \omega^4 & 0 & -\omega^3\sqrt{\phi} \\
    0 & 0 & \omega^4 & 0 & 0 \\
    \phi & 0 & 0 & 0 & -\omega^3\sqrt{\phi} \\
    0 & 0 & 0 & \frac{\omega^2}{\sqrt{\phi}} & \frac{-\omega^3\phi-\omega^2}{\sqrt{\phi}} \\
    \phi^{3/2} & -\omega & \omega^4 \sqrt{\phi} & \omega^2 & \omega^3-\omega^2 \\
  \end{pmatrix}, &
  \tilde{A}^{(23)} &= \begin{pmatrix}
    0 & 0 & \frac{\omega^4}{\phi} & 0 & 0 \\
    \phi & 0 & 0 & 0 & 0 \\
    0 & 0 & 0 & 0 & -\omega^3 \\
    0 & -\omega\sqrt{\phi} & 0 & \frac{\omega^2}{\sqrt{\phi}} & -\frac{\omega^2}{\sqrt{\phi}} \\
    0 & 0 & 0 & \frac{\omega^2}{\sqrt{\phi}} & -\frac{\omega^2}{\sqrt{\phi}} \\
    0 & -\omega\sqrt{\phi} & 0 & 0 & -\omega^3 \\
    \phi & -\omega\sqrt{\phi} & 0 & 0 & 0 \\
    \phi^{3/2} & \omega & \omega^4 & \omega^2 & -\omega^3\sqrt{\phi} -\omega^2 \\
  \end{pmatrix}, \notag \\[1ex]
  \tilde{A}^{(24)} &= \begin{pmatrix}
    0 & 0 & 0 & \frac{\omega^2}{\phi^{3/2}} & -\frac{\omega^2}{\phi^{3/2}} \\
    0 & 0 & 0 & 0 & -\omega^3 \\
    0 & -\omega & 0 & 0 & 0 \\
    \phi & 0 & \omega^4 \sqrt{\phi} & 0 & 0 \\
    \phi & 0 & 0 & 0 & 0 \\
    0 & -\omega & \omega^4 \sqrt{\phi} & 0 & 0 \\
    0 & 0 & \omega^4 \sqrt{\phi} & 0 & -\omega^3 \\
    \phi^{3/2} & -\omega\sqrt{\phi} & -\omega^4 & \frac{\omega^2}{\sqrt{\phi}} & \frac{-\omega^3\phi-\omega^2}{\sqrt{\phi}} \\
  \end{pmatrix}.
\end{align}

\end{document}